\documentclass[sigconf]{acmart}

\usepackage{booktabs} % For formal tables
\usepackage{amsfonts}
\usepackage[ruled]{algorithm2e}
\usepackage{multirow}
\usepackage[normalem]{ulem}

\renewcommand\footnotetextcopyrightpermission[1]{} % removes footnote with conference information in first column
\begin{document}
\title{Practical Constrained Optimization of Auction Mechanisms in E-Commerce Sponsored Search Advertising}

%\titlenote{Produces the permission block, and
%  copyright information}
%\subtitle{Extended Abstract}
%\subtitlenote{The full version of the author's guide is available as
%  \texttt{acmart.pdf} document}

\author{Gang Bai}
%\authornote{Dr.~Trovato insisted his name be first.}
\affiliation{%
  \institution{Alibaba Group}
  %\streetaddress{}
  %\city{Beijing}
  %\state{China}
  %\postcode{100083}
}
\email{baigang.bg@alibaba-inc.com}

\author{Zhihui Xie}
%\authornote{Dr.~Trovato insisted his name be first.}
\affiliation{%
  \institution{Tantan}
  %\streetaddress{}
  %\city{Beijing}
  %\state{China}
  %\postcode{100026}
}
\email{xiezhihui@p1.com}

\author{Liang Wang}
%\authornote{Dr.~Trovato insisted his name be first.}
\affiliation{%
  \institution{Alibaba Group}
  %\streetaddress{}
  %\city{Beijing}
  %\state{China}
  %\postcode{100083}
}
\email{liangbo.wl@alibaba-inc.com}

%\author{Who Who}
%\authornote{Dr.~Trovato insisted his name be first.}
%\affiliation{%
  %\institution{Alibaba Group}
  %\streetaddress{}
  %\city{Beijing}
  %\state{China}
  %\postcode{100083}
%}
%\email{who3@alibaba-inc.com}

\renewcommand{\shortauthors}{G. Bai et al}

\begin{abstract}

Sponsored search in E-commerce platforms such as Amazon, Taobao and Tmall provides sellers an effective way to reach potential buyers with most relevant purpose. In this paper, we study the auction mechanism optimization problem in sponsored search on Alibaba's mobile E-commerce platform. Besides generating revenue, we are supposed to maintain an efficient marketplace with plenty of quality users, guarantee a reasonable return on investment (ROI) for advertisers, and meanwhile,
facilitate a pleasant shopping experience for the users. These requirements essentially pose a constrained optimization problem. Directly optimizing over auction parameters yields a discontinuous, non-convex problem that denies effective solutions.
One of our major contribution is a practical convex optimization formulation of the original problem. We devise a novel re-parametrization of auction mechanism with discretized sets of representative instances. To construct the optimization
problem, we build an auction simulation system which estimates the resulted business indicators of the selected parameters by replaying the auctions recorded from real online requests. We summarized the experiments on real search traffics to analyze the effects of fidelity of auction simulation, the efficacy under various constraint targets and the influence of regularization. The experiment results show that with proper entropy regularization, we are able to maximize revenue while constraining other business indicators within given ranges.

\end{abstract}


%\keywords{ACM proceedings, \LaTeX, text tagging}

\maketitle

% begin of main body

\section{Introduction}

In this article, we present our work on auction mechanism optimization in the mobile sponsored
search engine of Alibaba's mobile E-commerce platform ({\it Taobao.com} and {\it Tmall.com}). In 2017, the platform powered millions of active advertisers to proactively reach hundreds of millions of unique potential buyers and effectively accomplish sales of goods worthy of hundreds of billions of RMB.

Sponsored search has been proved to be one of the most successful business model in online
digital advertising. For each user query, the sponsored search engine renders
relevant advertisements in addition to the main search results. The advertisers
bid on query keywords for their advertisements. For each advertisement showing
opportunity (an impression), the sponsored search engine selects a set of
advertisement candidates relevant to the search query, predicts their quality
scores such as the click-through-rate (CTR) and conversion-rate (CVR), allocates impression opportunities to advertisements using an auction mechanism 
and computes the clearing price for advertisers.
 
Generalized second price (GSP) mechanism is arguably the most widely used mechanism for the sponsored
search engines \cite{EOS2007,Aggarwal2006,Varian2007}, which ranks the advertisements by their bidding price and
quality score. The top ranked advertisements get the impressions and pay the
minimum price to maintain their ranking locations.

In the literature, most of the auction mechanisms focus on maximizing the
expected revenue in the Bayesian setting \cite{Myerson1981}, with variants on balancing efficiency and revenue using reserve prices \cite{Roberts2013,Ostrovsky2011}, or trading-off relevance and efficiency with an exponential weight of quality score\cite{Lahaie2011}. However, in the realistic
case of E-commerce, the auction mechanism should be optimized under many constraints including
the advertisers' budget, advertisement efficiency limits, etc. To make the platform revenue sustainable for long-term gain, we should
also take care of the factors like advertisers' return on investment (ROI,
quantitatively measured as the sale amount from the advertising cost) and users' satisfactory of the
search experience.

Moreover, most of the existing auction
mechanisms only work under very ideal environments, where the participants
are perfectly rational \cite{Tang17} and the click-through-rate of
advertisements are fixed according to the ad positions. However, these
assumptions are not true in industry search engines and the traffic characteristics
like user propensity, search queries changes dynamically over the days.

In Alibaba's mobile sponsored search platform, we conduct GSP auction in a virtual space (the ranking score). Hence, the key to improve the performance of our platform is to find the right ranking
score function. As mentioned above, our target is to maximize the revenue of the
platform while meeting the requirements like user experience and
advertisers' ROI. It is natural to formulate the auction optimization problem as a constrained optimization problem.

However, since the ranking function space is too large, it is impractical to explore the
performance of all the ranking functions online. To approximately gauge the outcome performance indicators of different ranking functions, 
we build a simulation system to replay the online auctions to generate virtual impressions and estimate expected user responses under a given ranking function.

To make the optimization problem practical, we introduce a discrete set of selected ranking functions as a novel representation of the auction mechanism. We re-parameterize the auction mechanism as the hitting probability of elements in the set. For each impression, one of the ranking functions is selected according to their probability to rank and price the candidate advertisements. The constrained optimization problem is to find the best hitting probability of the given set of ranking
functions. With this representation, we derive a convex optimization formulation of the problem.

\section{Problem and Formulation}

The auction process in E-commerce sponsored search platform can be formulated as follows: for each product search request with user query $q$, the search engine finds a set of advertisement candidates $Ad$ relevant to $q$ via broad match\cite{Yan2018} and estimate predicted CTR $\rho_a$, CVR $\rho'_a$ of each candidate using statistical models. Then the predicted CTR, CVR and bidding price $b_{aq}$ of the ad on the query keyword are mapped into a virtual space by evaluating a ranking score, and a GSP auction is conducted on the virtual space.

The ranking score, as the core of auctions in our platform, accounts for both expected efficiency\cite{Lahaie2011} and hidden cost\cite{Abrams2007}. It is principally defined as
\begin{equation}
\vartheta(\rho_a, b_{aq}, {\rho'}_a) = \underbrace{F_{\theta_e}(\rho_a)*b_{aq}}_{\text{expected efficiency}} + \underbrace{\hbar_{\theta_c}(\rho_a, {\rho'}_a)}_{\text{hidden cost term}} \label{eq:rankscore}
\end{equation}
where $\theta_e$ and $\theta_c$ are vectors of detailed parameters in the scoring functions. For simplicity, we denote the combined parameters as $\theta$. 
As a GSP auction in the space of the ranking score, the price for the $i$-th bidder in the ranking is determined as the infimum of bid that she can still keep her position
\begin{equation}
\mathrm{min}\, b  \textit{,  s.t. }  \vartheta(\rho_{a}^i, b, {\rho'}_a^i) \geq \vartheta(\rho_{a}^{i+1}, b_{aq}^{i+1}, {\rho'}_a^{i+1}) 
\end{equation}

Each component in $\theta$ weighs different input attributes of the function and influence the outcomes of ranking and pricing in the auctions differently, for example favoring higher bidding price or higher click probability. Due to business issues, we omit the detailed formulation of the components in \eqref{eq:rankscore}. This exclusion will not hinder the illustration of our methods though, since the method can be applied to various formulation of the ranking score function.

In practice, a context-aware mechanism which assigns different mechanisms on different traffic is usually used to better capture the distinct properties of search requests and ad candidate sets. In our case, search requests are designated into categories $\mathcal{C}$ using query information. Each category $c$ has a manually tuned parameter $\tilde{\theta}_c$ that effectively conducts the auctions.

For sustainable development of the platform, we should keep improving the satisfactory of users and advertisers while generating revenue for the platform. In our work, we regard the satisfactory of users as their engagement with the platform, {\it i.e.} the advertisement clicks and product purchases they make. For advertisers, we measure average cost for each user engagement through the advertisement as indicators of advertiser's ROI. We also take advertising PV coverage ratio
(ratio of ad impressions to total ad slots) as an important factor of users' search experience, as excessively displaying ads among search result is typically displeasing.

Different ranking function parameter $\theta$s have different outcomes of ranking and pricing, which lead to different user responses and eventually difference business metrics. We define business performance indicators of $\theta$ accordingly, including platform revenue, overall click-through rate (CTR), conversion rate (CVR) and ad PV coverage ratio (PVR), and advertisers' fulfillment of goal as cost per-click (CPC) and cost per-conversion/acquisition (CPA).

\subsection{Constrained Mechanism Optimization}

One may see this setup as a multi-objective optimization problem as proposed in \cite{Wang2012}. However, the objectives are indirectly and nonlinearly correlated and appropriately setting the preferences among each objective to meet a particular set of requirements is difficult.

Instead, complying with the general business objective, we formulate our business problem into a constrained optimization setup, which optimizes over auction parameters $\theta_1 \cdots \theta_{|\mathcal{C}|}$ for each category $c$. In detail, the goal is to find the best parameters of which the outcome of auctions maximizes the total revenue of the platform, while being feasible for the constraints on the metrics of CTR, PPC, CVR, CPA and PVR, etc. We denote targets of lower bounds of
CTR and CVR, lower and upper bounds of CPC, CPA and PVR as $\underline{\rho}$, $\underline{\rho}'$, $\underline{\pi}$, $\overline{\pi}$, $\underline{\pi}'$, $\overline{\pi}'$, $\underline{\kappa}$ and $\overline{\kappa}$, respectively. 

Directly solving the optimization problem is intractable since we are working on the second price auction mechanism, the outcomes are typically non-convex and even discontinuous with respect to the parameters $\theta$. It is fascinating and challenging to formulate the optimization problem into a sophisticated framework, such as convex optimization \cite{Boyd2004}.

In this article, we present a novel re-parametrization of ranking score function using a set of representative parameter instances to make the optimization problem practical. The instances in the set are selected by evenly discretizing each dimension of the parameter within a bounded-box centering at parameter $\tilde{\theta_c}$ of the original ranking function. In our experiments, the number of selected fixed parameter vectors $K$, as is the number of grids in the bounded-box, is 2025. 

This set of parameters $\{\theta_{c, j}$, $j \in 1 \cdots K\}$ for each request categories $ c \in \mathcal{C}$ provides a comprehensive range of different outcomes of auctions around that of the original $\tilde{\theta_c}$'s. By weighing the elements in the set, we are able to tune it to produce specific results.

To steer the outcome of the family of ranking score functions, we asign a probability of selecting the instance $\theta_{c,j}$ for category $c$, denoted as $x_{c,j}$. In application, the process of applying this family of mechanisms in sponsored search auction is described in Algorithm\ref{alg:online}. The re-parametrization with $x_{c,*}$ makes the constrained optimization problem practical.

We articulate the method of obtaining the best distribution $x_{c,j}$ over $\theta_{c,j}$ conforming to the business requirements. For each request $q_i$ of category $c$, the ranking function with parameter $\theta_{c,j}$ produces the corresponding ranking and pricing result, of which we denote the expected user response of click and purchase as $\rho_{i,j}$ and $\rho'_{i,j}$ and the price of click as $\pi_{i,j}$. With all the auction outcomes of each request with all the ranking function
instances, our problem is to find the particular probability values $x_{c,j}$ for each category and each ranking function of the category that maximizes expected revenue and meets the business requirements in expectation. Formally, the constrained optimization problem is materialized as:

\begin{align}
 \operatorname*{arg\,min}_{\mathbf{x}}\,\, & -\Sigma_{c}\,\Sigma_i \Sigma_j {\small \mathcal{I}\{q_i \in c \}}  x_{c,j} \rho_{i,j}\pi_{i,j} \label{prob:practical_obj}\\
 \text{s.t.   }\, & \Sigma_{c}\,\Sigma_i \Sigma_j {\small \mathcal{I}\{q_i \in c \}}  x_{c,j} \rho_{i,j} \geq \underline{\rho} \label{prob:ctr_distr_constraint} \\
    \underline{\pi} \leq & \frac{\Sigma_{c} \, \Sigma_i \Sigma_j {\small \mathcal{I}\{q_i \in c \}} x_{c,j} \rho_{i,j}\pi_{i,j}}{ \Sigma_{c}\,\Sigma_i \Sigma_j {\small \mathcal{I}\{q_i \in c \}} x_{c,j} \rho_{i,j}} \leq \overline{\pi}  \label{prob:ppc_distr_constraint} \\
    \underline{\kappa} \leq & \frac{\Sigma_c\,\Sigma_i \Sigma_j {\small \mathcal{I}\{q_i \in c \}}x_{c,j}{\small \mathcal{I}\{\text{ad}_i \text{ exists}\}}}{\Sigma_c\,\Sigma_i \Sigma_j {\small \mathcal{I}\{q_i \in c \}}x_{c,j} \mathbf{1}} \leq \overline{\kappa} \label{prob:pvr_distr_constraint} \\
   & \frac{\Sigma_{c} \, \Sigma_i \Sigma_j {\small \mathcal{I}\{q_i \in c \}} x_{c,j} \rho_{i,j}\rho'_{i,j} }{\Sigma_{c} \, \Sigma_i \Sigma_j {\small \mathcal{I}\{q_i \in c \}} x_{c,j} \rho_{i,j} } \geq \underline{\rho}' \label{prob:cvr_distr_constraint} \\
    \underline{\pi}' \leq & \frac{\Sigma_{c} \, \Sigma_i \Sigma_j {\small \mathcal{I}\{q_i \in c \}} x_{c,j} \rho_{i,j}\pi_{i,j} }{\Sigma_{c} \, \Sigma_i \Sigma_j {\small \mathcal{I}\{q_i \in c \}} x_{c,j} \rho_{i,j}\rho'_{i,j} } \leq \overline{\pi}' \label{prob:cpa_distr_constraint} \\
   & \sum_{j=1}^{K} x_{c, j} = 1, \text{  } \forall c \in \mathcal{C} \label{prob:sum_1} \\
   & x_{c, j} \geq 0, \text{  } \forall c \in \mathcal{C}, \forall j \in 1,\cdots,K \label{prob:prob}
\end{align}
where $\mathcal{I}\{x\} = 1$ if $x$ is true and 0 otherwise.

In this way, we simplify the complex continuous parameter optimization problem into a discrete K-armed bandits optimization problem. To construct this setup, we estimate the resulted business indicators just at selected discrete$\theta_{c, j}$ instead of every possible instance of $\theta$. When focusing on the fixed set of rules, we are able to improve the accuracy of the estimations. Also, compared with a solution of one single ranking function, a distribution of the fixed instances has a spectrum of much finer-grained outcome, since it is essentially a linear combination of the outcomes of the
ranking functions in the set. Also, when applied online in the stochastic environment, a distribution of multiple instances works more smoothly and robustly than a fixed one. 

\begin{algorithm}[t]
\SetAlgoNoLine
\LinesNumbered
\KwIn{$x, \Theta$}
\KwOut{A list of auction-winning ads.}
\ForEach{search request $q$}{
  {Determine the category $c$ of the request $q$}\;
  {Sample a parameter configuration {\small $j \sim Multinoulli(x_{c,1},\cdots,x_{c,K})$}}\;
  {Assign the parameters {\small $\theta \leftarrow \Theta_{c,j}$}}\;
  \ForEach{$Ad$ candidate in parallel}{
    {Estimate the click and conversion probability $\hat{\rho}$ and $\hat{\rho}'$}
    {Evaluate the ranking score $\vartheta_{\theta}(\hat{\rho}, b_{aq}, \hat{\rho}')$}\;
  }
  {Sort and filter $Ad$ candidates by the ranking score}\;
  {Calculate click price for top $N_{slot}$ ads in the sorted set}\;
  {Response with the result ad list}\;
}
\caption{Apply auction mechanism online}
\label{alg:online}
\end{algorithm}

\section{Implementation}

The key problem in constructing the problem setup in Eq.\eqref{prob:practical_obj}...\eqref{prob:prob} is to evaluate the coefficients which are determined by the auction outcome and stochastic user responses. Applying $\theta_{c,j}$ directly online to real traffics in short periods will merely result in observations with high variance, yet applying in longer period is unaffordable since it would seriously damage the performance of the platform when applying $\theta$ that fiercely violates the business constraints. We carry out a biased but smooth estimation via offline replay simulation.

After estimating all the business indicators and set up the coefficients in Eq.\eqref{prob:practical_obj}...\eqref{prob:prob}, it is fairly straightforward to
solve the problem using augmented Lagrangian method \cite{Birgin2014}. Also, an entropy regularization term is added to the optimization goal in Eq.\eqref{prob:practical_obj} to make the results robust to the error in the performance indicator computation by offline auction simulation.

\subsection{Offline Replay Simulation}

In this work, we approximate the auction outcomes under different mechanism parameters by replaying the recorded online auctions using the parameters. Compared with applying to real online traffic, the offline replay simulation is safe since it has no effect on online user experience and advertiser's ROI. More importantly, we are able to apply various ranking and pricing rules to the exactly same set of requests and ad candidates. Whereas online evaluation of rules are based on splitting of real traffics and so on different set of requests, which brings variance.

Replay logs consist of the request context, the query and the user profile, and the whole set of ads' information in auction including predicted CTR, CVR and bidding price. Under the hood, the ad serving module records each of the ad request and the algorithmic module records the CTR and CVR predictions for the request as well as bidding information of each ad candidate. The recorded data are then collected by ETL infrastructures, aggregated and joined by the request id and eventually stored on Alibaba cloud, where we implement the offline simulation pipeline. The raw log data is organized as a table partitioned by hours, which amounts to hundreds of TBs each day. 

Based on the replay log data, we are capable to apply any ranking and pricing function $\theta$ to a snapshot of the online traffic and obtain a set of winning ads for each ad slot. By implementing the ranking and filtering logics according to the counterparts in the online ad serving module, we make the offline simulation of the auctions produce the same ranking and pricing results as online's.

With outcomes of simulated ranking and pricing, we estimate the expected user response to further estimate the business indicator metrics. We utilize the CTR and CVR predictions $\hat{\rho}$ and $\hat{\rho}'$ to approximate the expectation of user click and conversion.

However, due to systematic bias caused by simplifying assumptions and variations in data distributions between training and serving time, the empirical mean of the predicted and the actual CTRs diverge, especially when the position effect of the ad slot escalates. This divergence brings bias to our estimations of metrics. We remedy this bias by statistically calibrating the predicted CTRs \cite{McMahan2013} during replay simulation and evaluation of metrics, so that their empirical mean matches the actual CTR. 

\begin{figure}
 \includegraphics[width=0.94\linewidth]{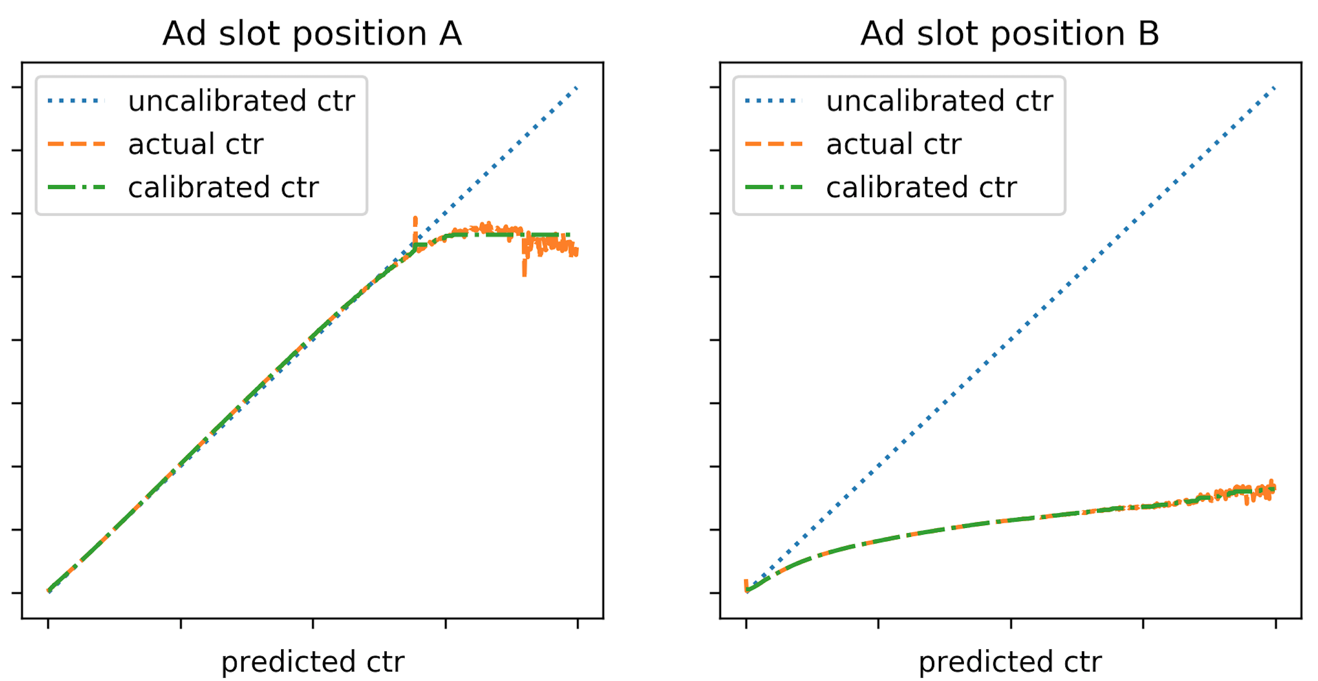}
 \caption{Calibration results of two ad slots.}
 \label{fig:calib}
\end{figure}

We learn calibration mappings $\varrho_p(\hat{\rho})$ for each advertisement slot position $p$ using isotonic regression\cite{Zadrozny2001}. $\varrho_p(\hat{\rho})$ is modeled as a piece-wise constant function, monotonically increasing with $\hat{\rho}$, which is a flexible approximator. The range of the input value $\hat{\rho}$ is split into $B$ intervals. For each interval $\tau$, within which the number of ad impressions is $\sigma_\tau$ and the actual CTR of these impressions is
$\check{\rho}_\tau$, we learn a constant factor $\omega_\tau$ as the calibrated
ctr corresponding to the interval of predicted CTR. The calibration learning task is formulated as:
\begin{align}
\operatorname*{arg\,min}_{\omega} & \sum_{\tau=1}^B \sigma_\tau|\omega_\tau-\check{\rho}_\tau|^2 \\
\text{subject to } & \omega_i \leq \omega_j \text{ if } i \leq j ,\, \forall i, j \in 1,\cdots, B 
\end{align}

We setup and solve the above learning task for each ad slot position. We prepare samples for calibration by aggregating advertisement serving logs. For each ad impression $i$ of slot position $p$, we extract the online predicted CTR $\hat{\rho}_i$, and label it is clicked or not. Then we aggregate the impression data by grouping by the interval bin $\tau$, and estimate the actual CTRs $\check{\rho}_\tau$ of samples that fall in each interval bin. 

Figure \ref{fig:calib} shows that calibration corrects the over-optimistically predicted CTRs and remediates the position bias in our platform.

With calibrated user response expectations, we calculate the business indicators on the whole dataset for each mechanism parameter $\theta_{c,j}$ in Eq.\eqref{prob:practical_obj}...\eqref{prob:prob}. With these steps, the construction of the problem setup is finished.

\subsection{Regularization with Entropy Bonus}

Solving the linear programming problem is fairly straightforward using sophisticated solutions to obtain a globally convergent solution. However, as it is a data-driven approach, the coefficients in the problem setup are approximated and the data distribution may also vary day after day. The optimal combination of mechanism parameters found from the simulated data is likely to be suboptimal.

To remedy this problem, we introduce some additional prior information via regularization to prevent overfitting to the replay simulated data. In our case, we borrow the idea of entropy regularization (entropy bonus) from reinforcement learning research \cite{Williams1991}, where entropy bonus was introduced to prevent convergence to a single choice of action and enforce exploration. 

We add this regularization term weighted by a hyper-parameter $\nu$ to the objective:

\begin{equation}
L(\mathbf{x}) = \sum_{c} \sum_j x_{c,j}(-\sum_i {\small \mathcal{I}\{q_i \in c \}} \rho_{i,j}\pi_{i,j} + \nu x_{c,j}\mathrm{ln}x_{c,j}) \label{eq:regularized_obj}
\end{equation}

With the objective Eq.\eqref{eq:regularized_obj} and constraints Eq.\eqref{prob:ctr_distr_constraint}...\eqref{prob:prob}, the problem setup is a linearly constrained convex optimization. Among many available methods, we use augmented Lagrangian for its simplicity.

\section{Experiments}

In this section, we present the experiments to measure the efficacy of the proposed method in auction mechanism optimization. We designed and launched several rounds of experiments on a fraction of the search traffic on the mobile search engine of the E-commerce platform. The baseline of the experiments is implemented the same as the main product version of the ranking score function, which consists of manually tuned parameters $\tilde{\theta}_c$ for each of the several thousands search
categories. From the experimental results, we studied the effects of various factors on the solution of the problem and summarized the motivations, the arrangements, the results and some analysis of our experiments.

For each experiment, we construct a particular setup of the constrained problem. We evaluate the estimated business indicators for all the fixed parameters $\theta_{c,j}$ via auction replay simulation on logged data from the past 14 days. In the same pipeline, we also simulate the baseline mechanism to estimate the baseline metrics of exactly the same requests. The constraint targets in the problem setup Eq.{\eqref{prob:ctr_distr_constraint}...\eqref{prob:prob}} are set by scaling up and down the estimated baseline business
indicators from simulation for the upper and lower bound targets of the constraints. 

We solve the constrained optimization using augmented Lagrangian method, in which the dual variables generally indicates the difficulty of satisfying each constraint. With some particular constraint targets, the residual of the constraint term may be large, making the solution infeasible. This is expected since the targets may be beyond the domain in which spans the outcome of all combinations of the ranking functions from valid distributions.

We launched the implementation of Algorithm \ref{alg:online} with each particular solution $x$ as well as the baseline onto online A/B test environment in our mobile search platform to handle 1\% of the real search traffics. Experiments are retained for at least 24 hours to sufficiently gauge the business performance indicators.

\subsection{Results and Analysis}

The experiments and results confirms the effectiveness of our approach and analyzes the effects of miscellaneous factors to it in optimizing auction mechanisms with constraints. Our approach generally meets the business requirements specified in constraint targets and maximizes the revenue. 

\subsubsection{Calibration in simulation.} 

We check the effects of calibration in offline simulation by comparing the results of problem setup with and without calibration. The experimental result shows that calibration helps significantly improve accuracy of the simulated metrics estimation and produce online results that approximately comply with the constraint targets, as illustrated in table \ref{table:with-calibration}. 

Without calibration, the simulated results tend to be over-optimistic on the overall user response, {\it i.e} CTR and CVR. This is because ads with high predicted CTRs and CVRs, which may be over-estimated, are more likely to win the auctions. Thus this experiment also express the influence of accuracy in estimations of the objective and the constraint targets in the problem setup.

\begin{table*}[h]
\caption{With and without calibration}
\label{table:with-calibration}
\begin{tabular}{cccccccccccc}
\toprule
\multicolumn{6}{c}{\textbf{problem setup configurations}} & \multicolumn{6}{c}{\textbf{result metrics}} \\
 \cmidrule(lr){1-6} \cmidrule(lr){7-12}
$\Delta$CTR & $\Delta$PPC & $\Delta$PVR & $\Delta$CVR & $\Delta$CPA & \textbf{calib} & $\Delta$REV& $\Delta$CTR & $\Delta$PPC & $\Delta$PVR & $\Delta$CVR & $\Delta$CPA \\
\midrule
\multirow{2}{*}{$\geq$2.5\%} & 
\multirow{2}{*}{-2\%$\sim$0} & 
\multirow{2}{*}{-0.5\%$\sim$0.5\%} & 
\multirow{2}{*}{-1\%$\sim$1\%} & 
\multirow{2}{*}{-1\%$\sim$1\%} &
  with & +0.06\% & +2.31\% & -2.07\% & -0.13\% & -0.75\% &-1.33\% \\
& & & & & without & 1.09\% & +0.89\% & +0.48\% & -0.28\% & -0.92\% & +1.12\% \\ 
\bottomrule
\end{tabular}
\end{table*}

\subsubsection{Efficacy with various constraint targets}

Most importantly, we want to examine the effectiveness of the proposed method in conforming with the specific requirements of the business performance while maximizing revenue.

We specify different business requirements on the performance by varying the upper and lower bounds in the constraints of the problem setup. With the solutions of the various setups, we examine how the resulted metrics correlate to the targets.

The various arrangement of targets and the corresponding online experiment results are illustrated in table \ref{tab:effective-targets}. The metrics in the experiment results generally follow the designated constraint targets, though miscue does exist due to the limit of the ranking function and exceedingly selected target values.

For CVR and CPA, we use simpler constraint targets because the expected number of conversions is less accurate, since it is calculated based on expected number of clicks, which is also approximated. 

\begin{table*}
\caption{Various constraint targets}
\label{tab:effective-targets}
\begin{tabular}{ccccccccccc}
\toprule
\multicolumn{5}{c}{\textbf{constraint target configurations}} & \multicolumn{6}{c}{\textbf{results}}\\
\cmidrule(lr){1-5} \cmidrule(lr){6-11}
$\Delta$CTR & $\Delta$PPC & $\Delta$PVR & $\Delta$CVR & $\Delta$CPA & $\Delta$REV &  $\Delta$CTR & $\Delta$PPC & $\Delta$PVR & $\Delta$CVR & $\Delta$CPA  \\
\midrule

$\geq$1\% & -1\%$\sim$0 &-0.5\%$\sim$0.5\% & $\geq$0 & $\leq$0 &

 +0.387\% & +1.37\% & -0.91\% & -0.06\% & +0.31\% & -1.22\% \\

$\geq$1.5\% & -1\%$\sim$0 &-0.5\%$\sim$0.5\% & $\geq$0 & $\leq$0 &

 +0.378\% & +1.35\% & -0.89\% & -0.07\% & +0.33\% & -1.21\% \\

$\geq$1.5\% & -2\%$\sim$0 &-0.5\%$\sim$0.5\% & $\geq$0 & $\leq$0 &

 +0.146\% & +1.69\% & -1.38\% & -0.14\% & +0.24\% & -1.62\% \\

$\geq$2\% & -2\%$\sim$0 &-0.5\%$\sim$0.5\% & $\geq$0 & $\leq$0 &

 -0.074\% & +1.96\% & -1.72\% & -0.28\% & +0.11\% & -1.83\% \\ 

$\geq$2.5\% & -2\%$\sim$0 &-0.5\%$\sim$0.5\% & $\geq$0 & $\leq$0 &

 -0.219\% & +2.28\% & -2.11\% & -0.34\% & -0.05\% & -2.06\% \\ 

$\geq$3\% &  -2\%$\sim$0 &-0.5\%$\sim$0.5\% & $\geq$0 & $\leq$0 & \multicolumn{6}{c}{infeasible constraints} \\
\bottomrule
\end{tabular}
\end{table*}

\subsubsection{Regularization}

We evaluate the effects of the entropy regularization term in the objective of the optimization problem Eq.\eqref{eq:regularized_obj}. In the experiment, we choose one single set of constraint, of which the targets are $\Delta$CTR$\geq$1.5\%, -1\%$\leq \Delta$PPC$\leq$0, -0.5\%$\leq \Delta$PVR$\leq$0.5\%, $\Delta$CVR$\geq$0 and $\Delta$CPA$\leq$0.

The experimental results are listed in table \ref{tab:regularization-term}. To articulate the effects of regularization, we also list the estimated business indicators of applying the solution to the replay simulation dataset in addition to the online experimental results.

Regularization term makes the offline simulated metrics suboptimal, but, when set appropriately, it improves the robust of the optimization result and performs better in the online traffic.

\begin{table*}[h]
\caption{Various regularization term}
\label{tab:regularization-term}
\begin{tabular}{ccccccccc}
\toprule
\multirow{3}{*}{$\nu$}  & \multicolumn{4}{c}{\textbf{simulated}} & \multicolumn{4}{c}{\textbf{online}}\\
\cmidrule(lr){2-5} \cmidrule(lr){6-9}
& $\Delta$CTR$_s$ & $\Delta$PPC$_s$ & $\Delta$PVR$_s$ & $\Delta$REV$_s$ & $\Delta$CTR$_o$ & $\Delta$PPC$_o$ & $\Delta$PVR$_o$ & $\Delta$REV$_o$ \\
\midrule
\textbf{0}       & +2.62\% & -1.35\% & +0.74\%  & +1.98\% &
 +0.78\% & -1.25\%  & +0.16\% &  -0.32\% \\
\textbf{1e-4} & +1.78\% & -1.02\% & +0.34\% & +1.08\% &
 +1.31\%  & -1.14\%  & -0.11\% & +0.045\% \\
\textbf{1e-2} & +1.53\% & -0.79\% & +0.11\% & +0.84\% &
 +1.35\% & -0.89\% & -0.07\% & +0.378\% \\
\textbf{1}       & +1.24\% & -0.65\% & -0.08\% & +0.501\%  &
 +1.13\% & -0.73\% & -0.15\% & +0.241\% \\
\bottomrule
\end{tabular}
\end{table*}

\section{Conclusion}

In this article, we present a constrained optimization formulation of the auction mechanism optimization problem for E-commerce sponsored search platform. We showed this formulation is practical and applicable with discretized parameterization of the auction mechanism. We illustrate the construct of the problem setup with calibrated offline simulation and the objective with entropy regularization to improve the robustness of the results. 

From the experimental results, we can conclude that our proposed methods do conform approximately with the specified constraint targets while maximizing the revenue. Another contribution of our work is the building of the auction simulation system configured to accommodate our experiments. Moreover, it is also extensible for other experiments that may require a replay of a large number of online auctions. We also would like to comment that it is obvious
that our method would cause some bidding behavior changes once enforced and the characteristic of online data distribution will drift away from the snapshot of offline data. Whereas this problem is resolved as models are updated in a daily basis while most of the active campaigns on our platform last for days or even weeks.

The effectiveness of the proposed method is crucially impacted by two factors. One is the accuracy of the offline simulation in estimating the business indicators. In the future work, we will work on incorporating online evaluated performance indicators into the problem setup to improve the accuracy. The other factor is the fix set of selected representative parameters $\theta_{c,j}$. The outcome of the auctions spans the space defined by the outcome of each
individual point in the set. A significant boost in the business performance requires judiciously selected candidates in the set and even a new design of the ranking score function, which is also an important future direction of this work.

% end of main body

%\bibliographystyle{ACM-Reference-Format}
\bibliographystyle{plain}
\bibliography{gopt}

\end{document}